\documentclass[fleqn,usenatbib, useAMS]{mnras}

\usepackage{graphicx}	
\usepackage{amsmath}

\usepackage{amssymb}	
\usepackage{multicol}        
\usepackage{bm}		
\usepackage{pdflscape}	
\usepackage{threeparttable}
\usepackage{multirow}
\usepackage{times}
\usepackage{enumitem}
\usepackage{url} 
\usepackage{color}
\usepackage{amsmath,bm}
\usepackage{float}
\usepackage{lineno}
\usepackage{longtable}
\usepackage[english]{babel}
\DeclareRobustCommand{\DE}[3]{#2}

\title[Probing the Galactic halo with RR Lyrae stars ]{Probing the Galactic Halo with RR Lyrae Stars -- V. Chemistry, Kinematics, and Dynamically Tagged Groups}

\author[Jonathan Cabrera Garcia et al.]{Jonathan Cabrera Garcia,$^{1}$
Timothy C.\ Beers$^{1}$
Yang Huang,$^{2,3}$
Xin-Yi Li,$^{4}$
Gaochao Liu,$^{5}$ 
Huawei Zhang,$^{6,7} $
\newauthor
Jihye Hong, $^{1} $
Young Sun Lee,$^{1, 8} $
Derek Shank,$^{1} $
 Dmitrii Gudin,$^{9}$
 Yutaka Hirai,$^{1,10}$\thanks{JSPS Research Fellow}
 and Dante Komater$^{1}$
\\
$^{1}$Department of Physics and Astronomy and JINA Center for the Evolution of the Elements, University of Notre Dame, Notre Dame, IN 46556, USA\\
$^{2}$School of Astronomy and Space Science, University of Chinese Academy of Sciences, Beijing 100049, P.\,R.\,China\\
$^{3}$Key Lab of Optical Astronomy, National Astronomical Observatories, Chinese Academy of Sciences, Beijing 100012, P.\,R.\,China\\
$^{4}$South-Western Institute for Astronomy Research, Yunnan University, Kunming 650500, P.\,R.\,China\\
$^{5}$Center for Astronomy and Space Sciences, China Three Gorges University, Yichang 443002, P.\,R.\,China\\
$^{6}$Department of Astronomy, School of Physics, Peking University, Beijing 100871, P.\,R.\,China\\
$^{7}$Kavli Institute for Astronomy and Astrophysics, Peking University, Beijing 100871, P.\,R.\,China\\
$^{8}$Department of Astronomy and Space Science, Chungnam National University, Daejeon 34134, Republic of Korea\\
$^{9}$Department of Mathematics, University of Maryland, College Park, MD 20742-4015 USA\\
$^{10}$Astronomical Institute, Tohoku University, 6-3 Aoba, Aramaki, Aoba-ku, Sendai, Miyagi 980-8578, Japan
}

\date{Accepted XXX. Received YYY; in original form ZZZ}

\pubyear{2023}

\begin{document}
\label{firstpage}
\pagerange{\pageref{firstpage}--\pageref{lastpage}}
\maketitle

\begin{abstract}
We employ a sample of 135,873 RR Lyrae stars (RRLs) with precise 
photometric-metallicity and distance estimates from the newly calibrated 
$P$--$\phi_{31}$--$R_{21}$--[Fe/H] and $Gaia$ $G$-band $P$--$R_{21}$--[Fe/H]  absolute
magnitude-metallicity relations of Li et al., combined with available 
proper motions from $Gaia$ EDR3, and 6955 systemic radial velocities from $Gaia$ DR3 and other sources, in order to explore the chemistry and kinematics of the halo of the Milky Way (MW).  This sample is ideally suited for characterization of the inner- and outer-halo populations of the stellar halo, free from the bias associated with spectroscopically selected probes, and for estimation of their relative contributions as a function of Galactocentric distance.  The results of a Gaussian Mixture-Model analysis of these contributions are broadly consistent with other  observational studies of the halo, and with expectations from recent MW simulation studies.  We apply the HDBSCAN clustering method to the specific energies and cylindrical actions ($E$, J$_{r}$, J$_{\phi}$, J$_{z}$), identifying 97 Dynamically Tagged Groups (DTGs) of RRLs, and explore their associations with recognized substructures of the MW.  The precise photometric-distance determinations ($\delta\, d/d < 5$\%), and the resulting high-quality determination of dynamical parameters, yield highly statistically significant (low) dispersions of [Fe/H] for the stellar members of the DTGs compared to random draws from the full sample, indicating that they share common star-formation and chemical histories, influenced by their birth environments. 

\end{abstract}

\section{Introduction}
\label{sec:Intro}

RR Lyrae stars (RRLs), which are old ($>$10\,Gyr), low-mass ($<1$\,M$_\odot$), generally metal-poor periodic pulsating variables stars located on the instability strip of the horizontal branch, have been used for decades to explore the nature of the Galactic halo of the Milky Way (MW).  Their utility as halo tracers is primarily driven by their relatively high luminosities ($M_V \sim$\,0.65; e.g., \citealt{Catelan2008,Muraveval2018}), and the fact that they are ``self selecting''; identified by their variability alone, and not based on spectroscopic follow-up of alternative probes, such as main-sequence turn-off and giant-branch stars, which often require complex target-selection criteria.  

The availability of well-sampled light curves from multiple photometric observations by the $Gaia$ project (\citealt{GaiaCollaboration2022}) has enabled large samples of RRLs to be assembled, from which accurate determinations of metallicity have been obtained \citep{Li2023}, based on refined calibrations of the relationships between [Fe/H] and parameters derived from light curves (e.g., the period and Fourier-decomposition parameters $\phi_{31}$ and $R_{21}$) for type RRab (fundamental-mode pulsating stars) and type RRc (first-overtone pulsating stars), with a scatter of about $\sigma_{\text{[Fe/H]}}$ = 0.2\,dex.  When combined with the newly calibrated $G$-band absolute magnitude-metallicity relations of \citet{Li2023}, such a sample provides halo tracers from the local neighbourhood out to Galactocentric distances on the order of 100 kpc. Once large-scale substructures (such as the Sagittarius Core and Stream), dwarf galaxy satellites of the MW, and globular clusters (GCs) are identified and removed, the remaining RRLs can be used to infer the relative contributions of the inner- and outer-halo populations as a function of distance.  

With the addition of proper motions and systemic radial-velocity measurements from $Gaia$ DR3 \citep{GaiaDR3}, RRLs also provide a powerful tool for examination of the nature of the halo of the MW based on their derived space motions. An analysis of the energy-action space, derived from positions and velocities, can reveal stars with similar dynamical properties, known as Dynamically Tagged Groups (DTGs), which are stars with common birthplaces that have been stripped from their parent structures (e.g., dwarf satellites and GCs). Previous analyses of DTGs (e.g., 
\citealt{Limberg2021,Shank2022a,Shank2022b}) and Chemo-Dynamically Tagged Groups (CDTGs; e.g., \citealt{Gudin2021,Shank2023,Zepeda2023}) have provided insight into the star-formation and chemical-evolution histories of their natal environments. For example, the inner-halo region of the Galaxy is thought to comprise a large amount of stellar debris from at least one ancient merger, the substructure Gaia-Sausage-Enceladus (GSE) \citep{Belokurov2018, Helmi2018}, and possibly others.  Clustering RRLs based on their energies and actions provides the opportunity to associate them with mergers and accreted components that occurred over the full assembly history of the MW \citep{Helmi1999}, making the importance of identifying DTGs crucial for this study.

In Paper I \citep{Liu2020} of this series, we constructed a catalogue of over 6000 RRLs with metallicity and systemic radial velocity measured from SDSS/SEGUE \citep{York2000, Yanny2009, Rockosi2022} and LAMOST \citep{LAMOST} spectra. 
In Paper II \citep{Wang2022}, we performed comprehensive studies of the global structure and substructures of the Galactic stellar halo, based on a sample of about 3000 RRLs. 
Paper III \citep{Liu2022} addressed the chemical and kinematical properties of the halo, based on a final sample of about 5000 RRLs. 
\citet{Li2023} used the spectroscopic catalogue of Paper I to calibrate the photometric metallicities and distances of RRLs from {\it Gaia} light curves.
This sample, together with RRLs with radial velocity (RV) measurements from Paper I and other sources, was therefore adopted to further halo studies in this series.   
In Paper IV (Zhang et al., submitted), the origin of the Oosterhoff dichotomy of RRLs is investigated based on samples of stars from large-scale surveys.
 
We now continue this work based on the much larger initial sample of over 135,000 RRLs from \citet{Li2023}, spanning from close to the Galactic Centre to over 100 kpc, derive changes in the metallicity distribution function (MDF) from the inner-halo to the outer-halo region, and analyse DTGs of RRLs in the halo and disk systems of the MW. The large numbers of RRLs in this catalogue with homogeneous metallicity and precise distance estimates can be used to draw a clearer multi-dimensional picture of the Galaxy. We carry out two approaches: (1) analysis of the change in the metallicity distribution function (MDF) of RRLs as  functions of radial distance from the Galactic Centre and vertical distance from the Galactic plane, to quantify the relative contributions of the inner- and outer-halo populations of the Galactic halo, and (2) an unsupervised dynamical clustering approach for the subset of RRLs with available full space motions, to identify DTGs that were once members of previous mergers and accreted components. 

This paper is organized as follows. In Section~\ref{sec:Data}, we briefly describe the adopted data, the observed properties of the full sample (magnitudes, distances, and metallicities), and compare with the subset of stars with available RVs.  We also describe the identification and removal of stars associated with large-scale structures, such as dwarf galaxies (including the Large and Small Magellanic Clouds), the Sagittarius Core (Sgr Core) and Sagittarius Stream (Sgr Stream, including the leading and trailing arms), GCs, and the removal of possible artefacts and potentially problematic stars (e.g., those with possibly compromised astrometry or blended sources, or with very high interstellar reddening), assembling a Cleaned Sample. Section~\ref{sec:StellarHalo} examines the distribution of derived [Fe/H] for stars in the Cleaned Sample, as functions of Galactocentric distance and distance from the Galactic plane. We perform a Gaussian Mixture-Model (GMM) analysis to determine the relative contributions of the inner- and outer-halo populations in each distance bin. Section~\ref{sec:Agama} briefly summarizes our determinations of dynamical parameters for the stars with available RVs. Section~\ref{sec:DTGs} describes the dynamical properties for this sub-sample, the identification of DTGs, and their association with known substructures in the halo and disk systems, GCs, and with previously identified dynamical groups. Section~\ref{sec:stats} describes the chemical structure of the DTGs, and an evaluation of their statistical significance.  Section~\ref{sec:summary} presents a summary of our results, along with future prospects.

\section{Data}
\label{sec:Data}

\begin{figure*}
\centering
\includegraphics[width=20cm]{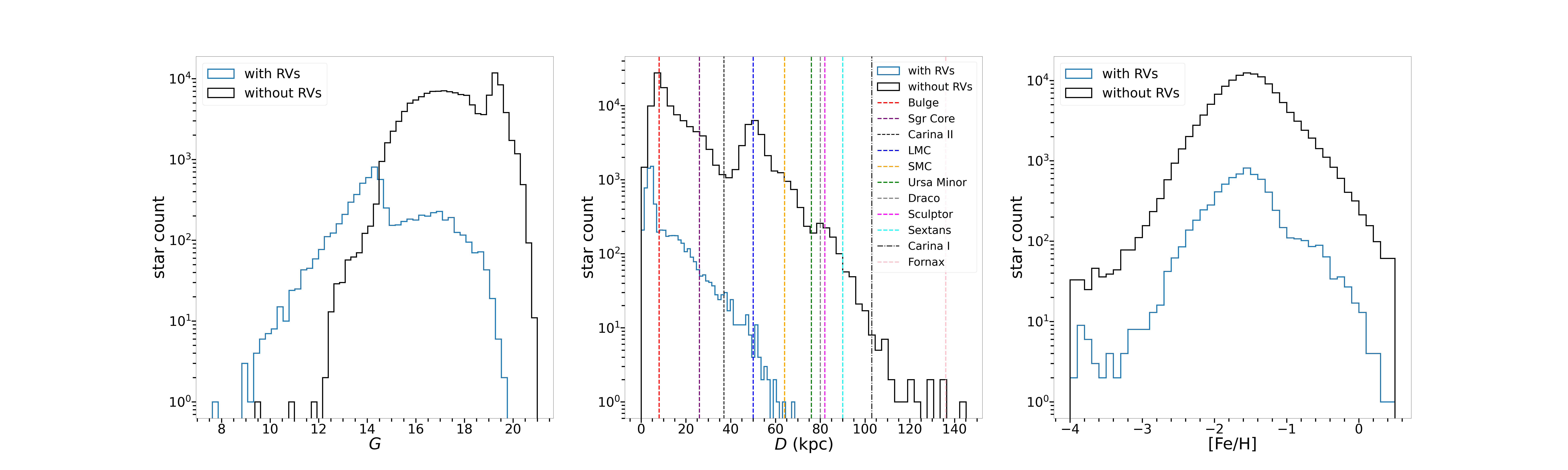}
\caption[Global parameters of RR Lyraes]{From left to right, the three panels display the $Gaia$ $G$ magnitude, the heliocentric distance $D$, and photometric-metallicity distributions of RRLs with and without RVs, respectively. The blue histograms correspond to the 6933 RRLs with available RVs, while the black histograms are for the 128,489 RRLs without RVs. The vertical dashed lines in the middle panel highlight known distances for Galactic Bulge, Sgr Core, the Magallenic Clouds (LMC and SMC), and dwarf galaxies: Carina I and II, Ursa Minor, Draco, Sculptor, Sextans, and Fornax. These distributions are based on the sub-sample of 135,422 RRLs with metallicities in the range $-4 < \text{[Fe/H]} \leq +0.5$.}
\label{fig:1}
\end{figure*}

\begin{figure*}
\centering
\includegraphics[width=20cm]{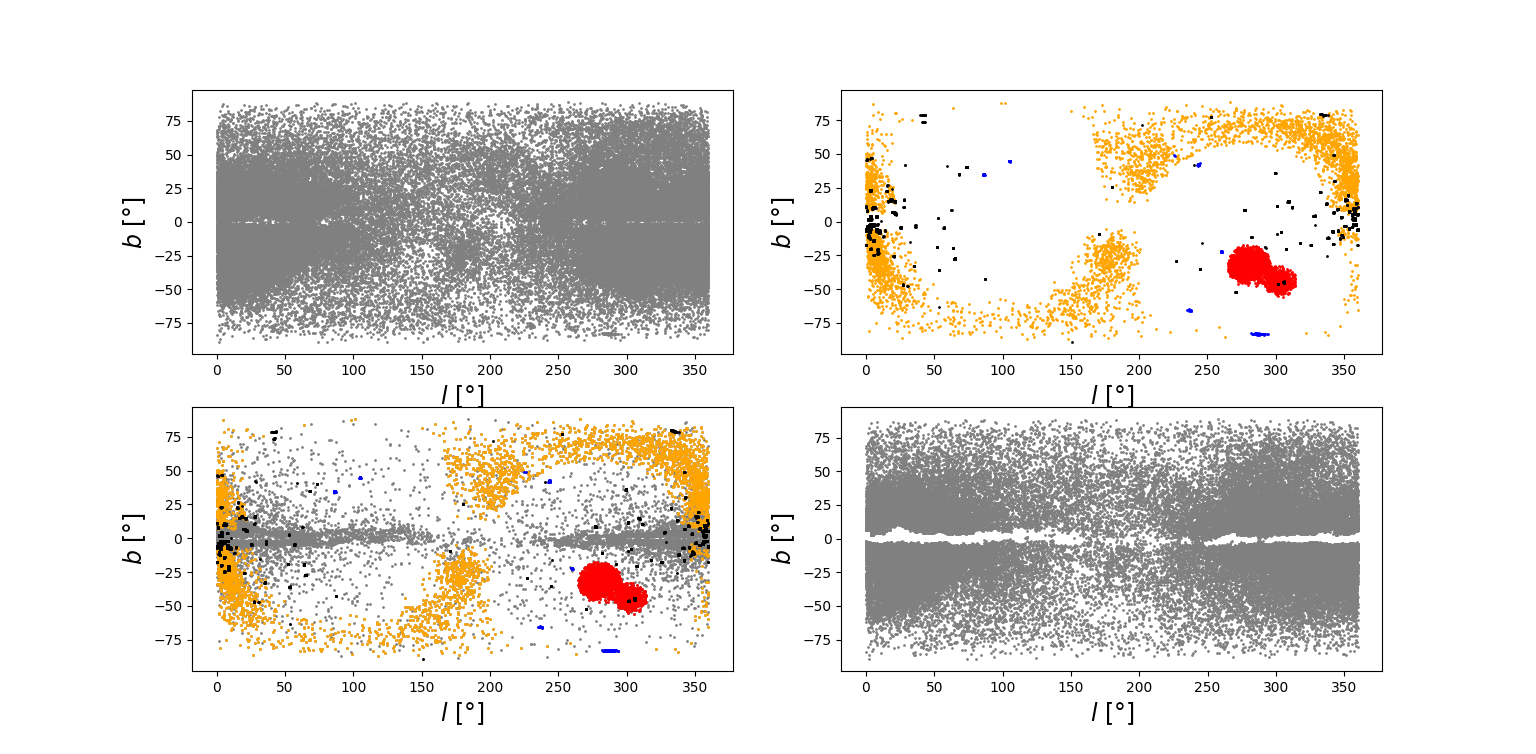}
\caption[Galactic coordinate Maps of data]{Top Left: Galactic coordinate map of the Initial Sample of 135,873 RRLs. Top Right: The sample of 32,356 RRLs identified as members of GCs (black), dwarf galaxies (blue), Sgr Stream and Sgr Core members (orange), and LMC or SMC members (red). Bottom Left: The sample of 56,975 RRLs that were removed from the Initial Sample if members are identified as MW substructures as indicated in the top-right panel or possible artefacts (gray), RRLs that might have compromised photometry or astrometry, as described in the text. Bottom right: The Cleaned Sample of 78,898 stars, after removing the sample of RRLs shown in the bottom-left panel. Note that the artefacts only correspond to the gray points in the 
bottom-left panel.}
\label{fig:2}
\end{figure*}

\begin{figure*}
\centering
\includegraphics[width=18cm]{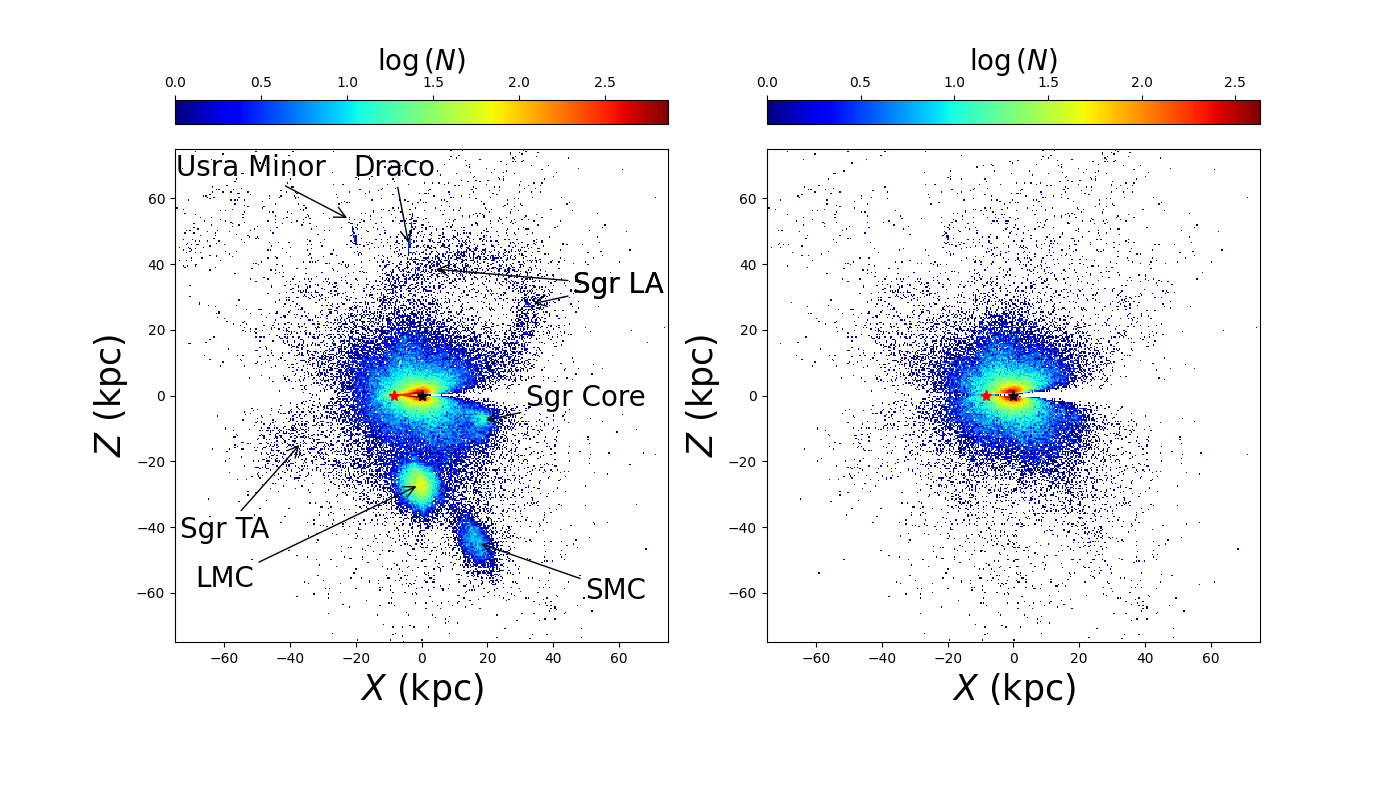}
\caption[Galactocentric Cartesian Map]{Galactocentric Cartesian map of the RRLs in the $Z$ vs. $X$ plane within 75 kpc. The left panel displays the sub-sample of 135,422 RRLs with metallicities in the range $-4 \leq \text{[Fe/H]} \leq +0.5$, along with labeled dwarf galaxies: Draco and Ursa Minor, the LMC, the SMC, and the Sgr Core, the Sgr leading arm (Sgr LA), and the Sgr trailing arm (Sgr TA). Note that there are more stars outside the ranges shown in the figure. The right panel shows the sample after trimming these labeled structures, the Cleaned Sample. The colour bar corresponds to the number of stars in each pixel, where the pixel size is 0.4 kpc by 0.4 kpc. The black star and red star represent the positions of the center of the Galaxy and Sun, respectively.}
\label{fig:3}
\end{figure*}

Our analysis is based on the catalogue of refined photometric estimates of metallicity ([Fe/H]) and distances of RRLs recently reported by 
\citet{Li2023}, which also includes proper motions from $Gaia$ EDR3 \citep{GaiaEDR3}, and systemic RVs for 6955 stars, where the vast majority of them are taken from $Gaia$ DR3 \citep{GaiaDR3}, and the rest are taken from \citet{Liu2020} and \citet{Clementini2022}. The Initial Sample of 135,873 RRLs consists of a total of 115,410 RRab and 20,463 RRc stars. Precise photometric-metallicity and distance estimates were derived by calibrating 
$P-\phi_{31}-R_{21}-\text{[Fe/H]}$, $P-R_{21}-\text{[Fe/H}]$, and $G-$band absolute magnitude relations, respectively; for details, see \citet{Li2023}. 

The global properties of the Initial Sample are presented in Fig. \ref{fig:1}, showing the distribution of the $G$-band magnitude (left panel), heliocentric distance $D$ (middle panel), and photometric metallicity $\text{[Fe/H]}$ (right panel) in this sample. Stars lacking available RVs are indicated with black histograms;
those with available RVs are shown in blue. In the middle panel, the vertical dashed lines indicate the distances of recognized structures, including the Galactic Bulge, the Sgr Core, the Large Magellanic Cloud (LMC) and Small Magellanic Cloud (SMC), and a number of dwarf galaxies.  Note that each panel only includes RRLs in the metallicity range $-4.0 < $[Fe/H] $\leq +0.5$. 

Fig. \ref{fig:2} presents maps of the Galactic coordinates ($l,b$) for our RRL stars. The panels show the Initial Sample (top left), the RRLs identified as belonging to recognized structures, including GCs, dwarf galaxies (the LMC, the SMC, and others) and the Sgr Stream and Sgr Core (top right), the RRLs that are removed as part of the trimming procedure described below (bottom left), and the Cleaned Sample of RRLs (bottom right) after these are excised, as described below.  

Heliocentric distances and Galactic coordinates are used to derive positions of RRLs in the Cartesian Galactocentric right-hand frame coordinate system, ($X$,$Y$,$Z$). Fig. \ref{fig:3} shows a map of the Initial Sample of RRLs within 75 kpc from the Galactic Centre. The left panel labels the recognized structures. For the first part of our analysis below, we discard potential members of GCs, the LMC, the SMC, and other dwarf galaxies, the Sgr Stream and Sgr Core, and potential artefacts, as described below, in order to focus on the 
inner- and outer-halo components of the Galaxy. We retain these structures (but remove the artefacts) for the dynamical analysis.

\subsection{Trimming the Initial Sample}
\label{subsec:data_sub1}

\subsubsection{Removal of globular clusters and dwarf galaxies}
\label{subsec:data_sub1_sub1}
Using the catalogue of GCs from \citet{Harris2010} with known equatorial positions $(\alpha, \delta)$ and their corresponding half-light radii $r_{h}$, RRLs are removed that are located within ten times the half-light radius for each GC. As a result, 3085 RRLs are discarded from the Initial Sample. For the dwarf spheroidal satellites, RRLs are removed within an angular distance of $0.5^{\circ}$ from the center of Carina I, Carina II, and Ursa Minor, $0.3^{\circ}$ from Leo I and Leo II, and $1^{\circ}$ from Sculptor, Fornax, Draco, and Sextans. The centers of these dwarf galaxies are quoted in Table C.2 of 
\citet{GaiaCollaboration2018}. In the process, a total of 530 RRLs are removed from the Initial Sample.  

\subsubsection{Removal of the Magellanic Clouds}
\label{subsec:data_sub1_sub2}

The LMC and SMC are the largest nearby satellites of the Galaxy, and can be seen clearly in Fig. \ref{fig:3}. To remove the likely RRL members of the Magallenic Clouds, we first consider stars that fall within $16^{\circ}$ and $12^{\circ}$ from the LMC and SMC, respectively. Secondly, the subset of those stars with proper motions relative to the LMC or SMC less than $\pm 5$ mas yr$^{-1}$ with respect to the values reported in \citet{GaiaCollaboration2018} are kept. Next, we require that RRLs have $G > 18.5$, based on the expected apparent magnitude of the RRL stars at the distances of the LMC and SMC \citep{Belokurov2017}. Finally, stars are selected within 5$^{\circ}$ from the LMC and SMC centers and with distance modulii $\mu_{\text{LMC}} > 17.5$ or $\mu_{\text{SMC}} > 18.0$ to treat the dense central region of the Magallenic Clouds. In this process, a total of 22,869 RRLs are removed from the Initial Sample.

\subsubsection{Removal of the Sagittarius Stream and Sagittarius Core}
\label{subsec:data_sub1_sub3}
Discovered in 1994 \citep{Ibata1994}, the Sgr dwarf galaxy is an ancient relic, crucial for exploring the gravitational potential of the MW and for understanding how its leading and trailing stream arms are influenced by tidal disruption. As seen in Fig. \ref{fig:3} 
(left panel), the Sgr Core is a relatively dense region located at around ($X=20 $, $Z=-5 $) kpc. Using the sample of RRL Sgr Stream and Sgr Core candidates from 
\citet{Ramos2020}, we cross-match with our Initial Sample via $Gaia$ DR3 source IDs. A total of 5872 RRLs are identified as potential Sgr Stream and 
Sgr Core members, and removed from the Initial Sample. 

\subsubsection{Removal of artefacts}
\label{subsec:data_sub1_sub4}
After the removal of potential members of substructures, artefacts are also considered in the trimming process. These artefacts are identified based on the following three criterion:

\begin{enumerate}[label=(\roman*)]
\item RUWE $>$ 1.2
\item BRE $>$ 1.5
\item $E(B-V) > $ 0.8,
\end{enumerate}
where RUWE is the renormalized\_unit\_weight\_error, BRE is the phot\_bp\_rp\_excess\_factor, representing the ratio between the combined flux in $Gaia$ $BP$- and $RP$- bands ($G_{\text{BP}}$ and $G_{\text{RP}}$) and the total flux in the $G$-band, and $E(B-V)$ is the reddening. Criterion (i) is applied to remove sources whose astrometric solutions are not well-represented by the single-star five-parameter model, as described in \citet{Lindegren2018}. Criterion (ii) is applied to identify sources that are considered blends due to their high BRE values (see \citealt{Evans2018}). Criterion (iii) is applied to remove stars in regions with very high reddening. The number of artefacts in the Initial Sample that meet one or 
more of the three criteria is 24,619, which are excised from the sample. 

\subsection{The Cleaned Sample}
\label{subsec:data_sub2}

We define the Cleaned Sample of RRLs by application of the trimming procedure described above, resulting in a total of 78,898 stars (see the bottom-right panel of Fig. \ref{fig:2} and the right panel of Fig. \ref{fig:3}).  We considered whether we could recover some of the trimmed stars by relaxing one or more of the criteria used. For example, criterion (i), which guards against poor astrometry, is not a concern for RRLs with photometric-distance estimates, and for which proper motions do not enter the calculation, and rejection on the basis of criterion (ii), which may be impacted by the nature of the flux distribution within the $Gaia$ photometric bands for strongly variable sources such as RRLs.  However, as described below, when we attempt to better isolate the halo populations of the MW by imposing a selection on distance from the Galactic plane of $|Z| > 3$ kpc, we find that most of the ``recovered" stars are eliminated by this additional cut, so this attempt was abandoned.  

For our present analysis, we also decided to only include stars with photometric-metallicity estimates in the range $-4.0 <$ [Fe/H] $\leq $ +0.5, on the grounds that the estimated [Fe/H] outside this range may be spurious, and require spectroscopic follow-up to be used with confidence. This cut only removed a relatively small number of stars, resulting in a total of 78,740 RRLs in the 
Cleaned Sample. 

\begin{figure*}
\centering
\includegraphics[width=17cm]{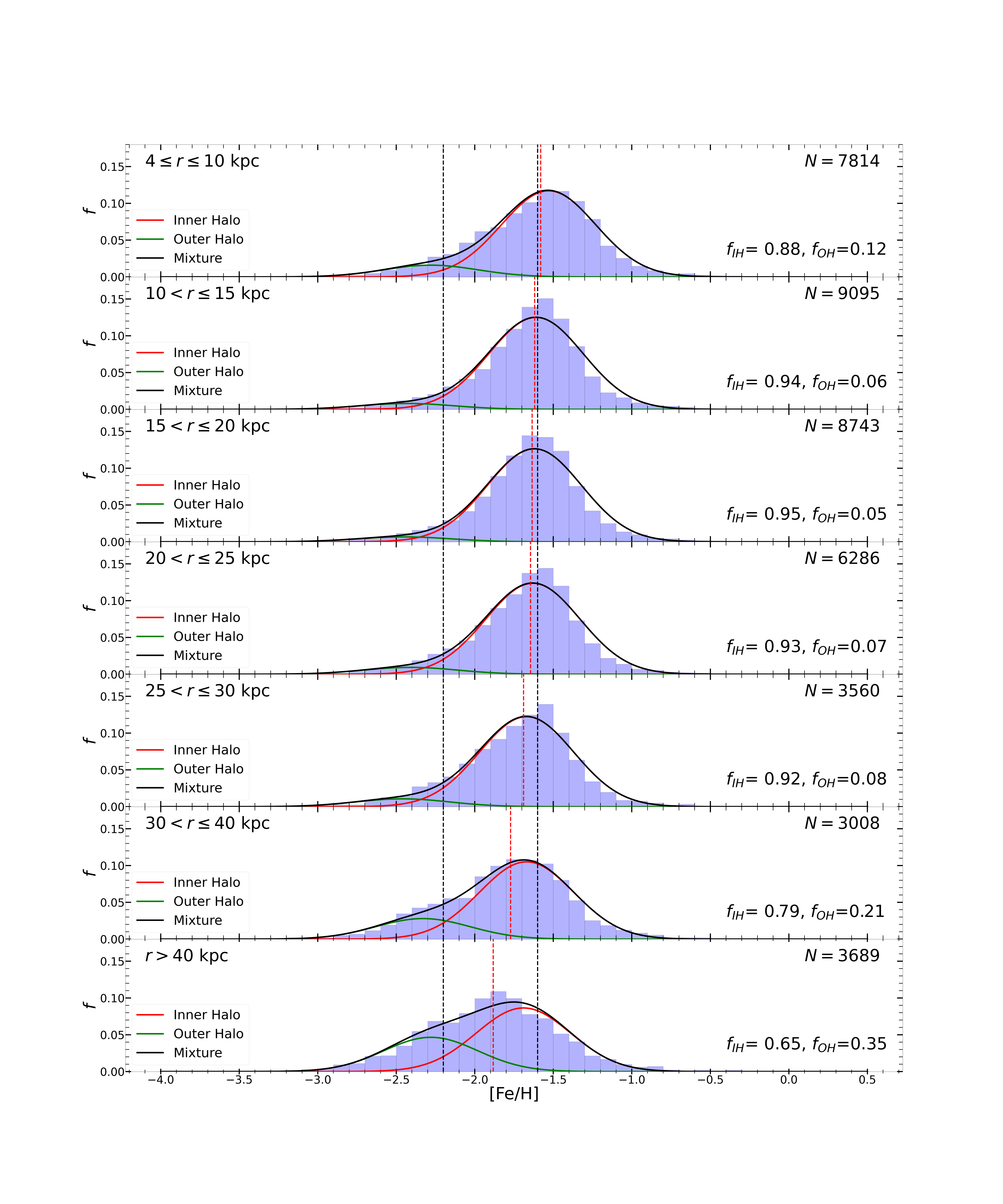}
\caption[Metallicity distribution functions of RRLs in the Stellar Halo]{The 
Cleaned Sample of  42,195 RRLs with $-4.0 < \text{[Fe/H]} \leq +0.5$ (x-axis) and $|Z| > 3$ kpc, divided into seven radial regions in Galactocentric distance within the halo. The y-axis displays the fraction of RRLs in a particular [Fe/H] bin with widths of 0.1 dex. The upper-left corner of each panel lists the region in the 
stellar halo considered, and the upper-right corner shows the number of RRLs in that distance bin. The two black dashed vertical lines are set at [Fe/H] = $-2.2$ and [Fe/H] = $-1.6$, as determined for the inner-halo and outer-halo populations by \citet{Carollo2007,Carollo2010} and \citet{Beers2012} on the basis of kinematic analyses of metal-poor stars from SDSS/SEGUE in a limited local volume. The red dashed line indicates the median metallicity of stars in each radial bin.  The transition in median metallicity with increasing radial distance is clearly evident, in particular beyond 20 to 25 kpc.  The solid red and green lines indicate the best fits to presumed inner-halo and outer-halo components obtained by a Gaussian mixture-model analysis; the solid black line indicates the sum of these components. The lower-right corner shows the fractions of each component.}
\label{fig:4}
\end{figure*}

\begin{figure*}
\centering
\includegraphics[width=17cm]{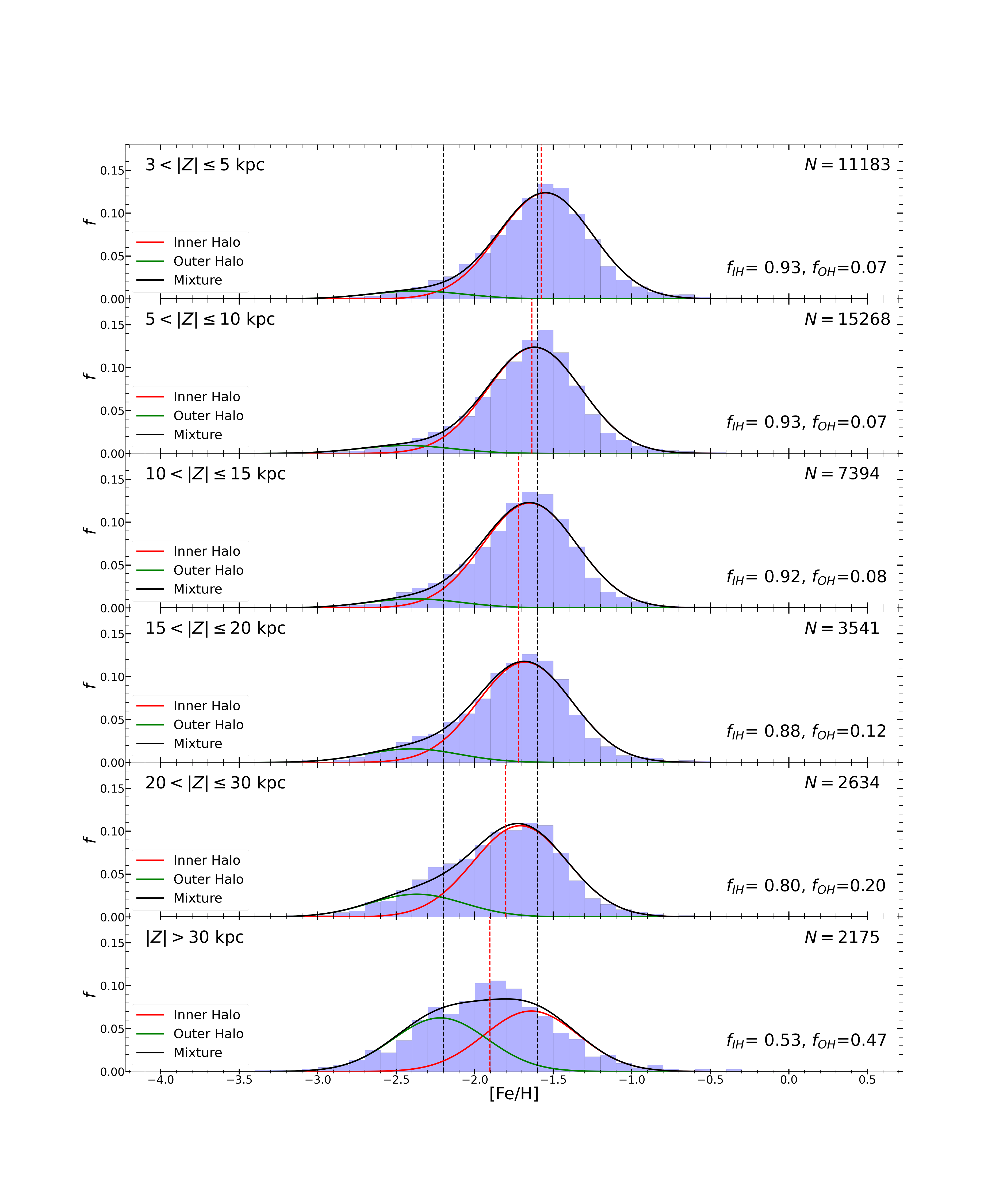}
\caption[Metallicity distribution of RRLs in the Stellar Halo]{Same as Fig. \ref{fig:4}, but divided into six vertical regions of distance from the Galactic plane within the halo.}
\label{fig:5}
\end{figure*}

\section{[Fe/H] distribution of RR Lyraes in the Stellar Halo}
\label{sec:StellarHalo}

Fig. \ref{fig:4} shows the MDFs for different 
radial regions in Galactocentric distance in the stellar halo within the metallicity range $-4.0 <$ [Fe/H] $\leq $ +0.5, after applying the additional $|Z| > 3$ kpc cut to the Cleaned Sample, which left a total of 42,195 stars.  Superposed on each panel are two black dashed vertical lines, which demarcate the metallicity peaks of the inner-halo and outer-halo stellar populations, as derived by \citet{Carollo2007,Carollo2010} and \citet{Beers2012} on the basis of kinematic analyses of metal-poor stars from SDSS/SEGUE in a limited local volume.  The red dashed line in each panel indicates the position of the median metallicity in each radial bin for comparison.
From inspection of this figure, it is clear that there is a general trend for decreasing [Fe/H] with increasing Galactocentric distance, as has been suggested by numerous previous analyses (e.g., \citealt{Carollo2007, Carollo2010, deJong2010, Beers2012, An2020, Dietz2020, An2021}). It is also apparent that the fractions of stars near the black dashed line marking the peak metallicity of the inner-halo population ([Fe/H] = $-1.6)$, while initially roughly constant, begins to drop beyond 20-25 kpc, while the fractions of stars near the peak metallicity of the outer-halo population ([Fe/H] = $-2.2$) increases beyond this radius. 

Fig. \ref{fig:5} shows the MDFs for different 
vertical regions from the Galactic plane in the stellar halo for this same sample.  The black dashed lines and red dashed line are as in Fig. \ref{fig:4}. 
From inspection of this figure, a general trend for decreasing [Fe/H] with increasing distance from the Galactic plane is clearly observed, as inferred by 
\citet{Carollo2010} based on the kinematics of local halo stars. The fractions of stars near the black dashed line marking the peak metallicity of the inner-halo population, while initially roughly constant, begins to drop beyond 10-15 kpc, while the fractions of stars near the peak metallicity of the outer-halo population increases beyond this distance. 

In order to gain more insight about the nature of the stellar populations in the Galactic halo, and their relative contributions as a function of Galactocentric distance and distance from the Galactic plane, we have carried out a GMM analysis, as described below. 

\subsection{Gaussian Mixture-Model Analysis}
\label{subsec:StellaHalo_sub1}

Similar to \citet{Liu2022}, we employ two Gaussian components to
model the contribution from the inner-halo (IH) and outer-halo (OH) populations, respectively. 
The metallicities  are assumed to follow a Gaussian distribution with means $\rm[Fe/H]_{\rm IH}$ / $\rm[Fe/H]_{\rm OH}$ and dispersions $\sigma_{\rm IH}$ /$\sigma_{\rm OH}$. 
The fractions of IH and OH stars are represented by $f_{\rm IH}$ and $f_{\rm OH}$, respectively.  In this way, for the model parameters $\Theta =$\,\{$\rm[Fe/H]_{\rm IH}$,\,$\sigma_{\rm IH}$,\,$f_{\rm IH}$,\,$\rm[Fe/H]_{\rm OH}$,\,$\sigma_{\rm OH}$\}, the likelihood of observing the $i$th sample star with metallicity ${\rm [Fe/H]}_{i}$ is given by:

\begin{equation}
L_{i} (\rm[Fe/H]_{i}|\Theta) = \it{f}_{\rm IH}\it{P}_{\rm IH} ({\rm [Fe/H]}_{i}) + (\rm{1} - \it{f}_{\rm IH})\it{P}_{\rm OH}({\rm [Fe/H]}_{i})\text{.} 
\label{eq1}
\end{equation} 

The probability $P_{\rm IH} ({\rm [Fe/H]}_{i})$ of a star with measured metallicity [Fe/H]$_{i}$ belonging to the inner-halo population can be calculated from:

\begin{equation}
P_{\rm IH} ({\rm [Fe/H]}_{i}) = \frac{1}{\sqrt{2\pi}\sigma_{\rm IH}}\exp{\left[-\frac{1}{2}\left(\frac{{\rm [Fe/H]}_{i}-\rm[Fe/H]_{\rm IH}}{\sigma_{\rm IH}}\right)^2\right]}\text{.} 
\label{eq2}
\end{equation}
       
  The probability $P_{\rm OH}({\rm [Fe/H]_{i}})$ of a star with measured metallicity $\rm{[Fe/H]}_{i}$  belonging to the outer-halo population can be obtained from:
  
\begin{equation}
P_{\rm OH}({\rm [Fe/H]}_{i}) = \frac{1}{\sqrt{2\pi}\sigma_{\rm OH}}\exp{\left [-\frac{1}{2}\left (\frac{{\rm [Fe/H]}_{i}-\rm[Fe/H]_{\rm OH}}{\sigma_{\rm OH}}\right )^2\right ]}\text{.} 
\end{equation}

The likelihood of a specific bin $j$ is calculated by multiplying the function of equation \ref{eq1} for a total of $N_j$ stars located within bin $j$:

\begin{equation}
\centering
L = \prod_{i=1}^{N_j} L_i\text{.} 
\end{equation}

The posterior distribution of the model parameters is obtained by:

\begin{equation}
p (\Theta|{\boldsymbol {O}}) \propto L({\boldsymbol {O}}|\Theta)I(\Theta)\text{,}
\end{equation}

\noindent where \textbf{$O$} represents the observables, i.e., [Fe/H]$_{i}$,
and $I (\Theta) $ represents the priors of the model parameters.
In this study, a Bayesian Markov Chain Monte Carlo (MCMC) technique is used to obtain the posterior probability distributions of the model parameters. The best-fit values of the model parameters are then delivered by the marginalized posterior probability distributions shown in Figs. \ref{fig:4} and \ref{fig:5} and summarized in Tables \ref{tab:1} and \ref{tab:2} for the  radial and vertical bins, respectively.

\begin{table*}
\caption{GMM fitting results for radial bins in Galactocentric distance.  [Fe/H]$_{\rm IH}$, $f_{\rm IH}$, $\rm{[Fe/H]}_{\rm OH}$, and $f_{\rm OH}$ indicate the fitted mean values of metallicity and fractions of the inner-halo and outer-halo components, respectively. }
\centering
\begin{tabular}{c|c|c|c|c|c|}
\hline
\hline
 Radial bins (kpc) & Number  &  [Fe/H]$_{\rm IH}$   &$f_{\rm IH}$ &$\rm{[Fe/H]}_{\rm OH}$ &$f_{\rm OH}$  \\
\hline
$4 < r \leq10$&7803  &$-1.53$&$0.88$&$-2.28$&$0.12$\\
$10 < r \leq15$&9085 &$-1.61$&$0.94$&$-2.43$&$0.06$\\
$15 < r \leq 20$&8736&$-1.62$&$0.95$&$-2.45$&$0.05$\\
$20 < r \leq 25$&6280&$-1.63$&$0.93$&$-2.41$&$0.07$\\
$25 < r \leq 30$&3551&$-1.67$&$0.92$&$-2.46$&$0.08$\\
$30 < r \leq 40$&3004&$-1.67$&$0.79$&$-2.33$&$0.21$\\
$r > 40$           &3681&$-1.69$&$0.65$&$-2.28$&$0.35$\\
\hline
\end{tabular}
\label{tab:1}
\end{table*}

 \begin{table*}
\caption{GMM fitting results for vertical bins in height above the Galactic plane (centered on the Sun). [Fe/H]$_{\rm IH}$, $f_{\rm IH}$, $\rm{[Fe/H]}_{\rm OH}$, and $f_{\rm OH}$ indicate the fitted mean values of metallicity and fractions of the inner-halo and outer-halo components, respectively. }
\centering
\begin{tabular}{c|c|c|c|c|c|}
\hline
\hline
 Vertical bins (kpc)  & Number & [Fe/H]$_{\rm IH}$   &$f_{\rm IH}$ &$\rm{[Fe/H]}_{\rm OH}$&$f_{\rm OH}$    \\
\hline
$3 < |Z| \leq 5$  &11183&$-1.55$&$0.93$&$-2.37$&$0.07$\\
$5 < |Z| \leq 10$ &15268&$-1.62$&$0.93$&$-2.44$&$0.07$\\
$10 < |Z| \leq 15$&7394&$-1.65$&$0.92$&$-2.38$&$0.08$\\
$15 < |Z| \leq 20$&3541&$-1.68$&$0.88$&$-2.40$&$0.12$\\
$20 < |Z| \leq 30$&2634&$-1.71$&$0.80$&$-2.37$&$0.20$\\
$|Z| >  30$       &2175&$-1.64$&$0.53$&$-2.22$&$0.47$\\
\hline
\end{tabular}
\label{tab:2}
\end{table*}

\subsection{The Dual-Halo Interpretation}
\label{subsec:StellarHalo_sub2}

From inspection of the dual-halo model fits shown in Figs. \ref{fig:4} and \ref{fig:5}, this simple picture does a credible job capturing the change in the MDF for the radial and vertical regions considered.  The initial parameter guesses (the peaks of the MDFs for these components obtained by \citealt{Carollo2010}), based on a kinematic analysis of local metal-poor stars, leads to fitted mean metallicities that agree with these to within 0.1 to 0.2 dex across all regions.  There remain deviations between the observations and the mixture model in some cases, but none sufficiently large that introduction of additional components appears to be required. It should be kept in mind that our procedure for identifying and removing clear substructure from the data set described above may well be imperfect, so some deviations between the mixture-model predictions and the observations might be associated with this ``residual substructure."   

There originally existed some controversy about whether or not target-selection biases or other systematic errors (such as in distance or proper-motion estimates) in the \citet{Carollo2007} and \citet{Carollo2010} papers that originally presented kinematic evidence for the existence of a dual halo, which were largely refuted by \citet{Beers2012}. In Paper III \citep{Liu2022}, on order of 5000 RRLs were studied kinematically and chemically, revealing that the Galactic stellar halo comprises an IH and OH population, changing their relative importance at a Galactocentric radius of 30 kpc. Our demonstration here that, based on a very large, and essentially bias-free selection of \textit{in-situ} halo RRLs, the Galactic halo comprises at least two stellar populations of likely different origin, removes any remaining doubt. Numerous other studies of halo stars, using alternative samples of tracers and methods, have also appeared in the past decade that bolster the veracity of the existence of a dual halo \citep[e.g.,][] {deJong2010, Xue2011, Carollo2012, Carollo2016, 
Hattori2013, Lee2013, Allende_Prieto2014, Fernandez-Alvar2016, Das2016, Battaglia2017, Helmi2017, Helmi2018, Belokurov2018, Yoon2018, Lee2019, Whitten2019, An2021, Liu2022}. 

Numerical simulations of MW-like galaxies have also provided support for a dual-halo interpretation \citep[e.g.,][]{Zolotov2009, Font2011, Font2020, McCarthy2012, Tissera2014, Cooper2015, Pillepich2015, Rey2022}.  For example, \citet{Zolotov2009} found that the inner regions ($r<20$ kpc) of their simulated stellar haloes are composed of both \textit{in-situ} and accreted components, while stars in the outer region originated from accreted satellites. More recently, the stellar haloes in the \textsc{artemis} simulations have shown that the stellar-density profiles can be fitted with broken power laws \citep{Font2020}. Their results suggest that the break radii (typically $\approx$ 20--40 kpc) correspond to the transition between \textit{in-situ} formation and accreted origin. With a semi-empirical approach, \citet{Rey2022} pointed out that a prominent last major merger results in a steeper radial profile for the accreted components. These simulations support that stellar haloes in MW-like galaxies tend to have dual haloes. These dual haloes are composed of inner haloes formed by \textit{in-situ} and accreted components and outer haloes dominated by accreted components.

\section{Dynamical Parameters Determined with \texttt{AGAMA}}
\label{sec:Agama}
From the Initial Sample (135,873 RRLs), there are 6955 RRLs with RVs, 
and 6932 RRLs that also are within the metallicity range $-4.0 < \text{[Fe/H]} \leq +0.5$. After application of the artefact-removal conditions described in Section~\ref{subsec:data_sub1_sub4} above, our Final Sample of RRLs with RVs contains a total of 5355 stars. We employ this subset of stars in order to estimate dynamical parameters as summarized here.

\begin{table*}
\centering
\caption{Identified DTGs, numbers of stars, confidence, mean abundance and dispersion, and associations.}
\begin{tabular*}{1.00\textwidth}{cccccc}
\hline
\hline
DTG & $N$ stars & Confidence & $\mu \pm \sigma$([Fe/H]) & Associations\\
\hline
1&38&72.0\%&$-1.69\pm0.33$&Helmi Stream
, FW22:Group73, EV21:NGC 5272 (M 3), FW22:Group71\\
&&&&FW22:Group74, FW22:Group70, FW22:Group72, ZY20a:DTG-3\\
&&&&FW22:Group69, FW22:Group75, DS22b:DTG-42, DG21:CDTG-15, GL21:DTG-3\\
2&27&38.0\%&$-1.56\pm0.21$&GSE, FW22:Group16, FW22:Group10, FW22:Group22\\
&&&&FW22:Group30, FW22:Group51, FW22:Group5, FW22:Group52\\
&&&&FW22:Group12, SL22:3, GM17:Comoving, GC21:Sausage\\
3&26&37.2\%&$-1.73\pm0.26$&GSE, FW22:Group3, FW22:Group11, FW22:Group6\\
&&&&DS22a:DTG-10, KH22:DTC-2, GL21:DTG-23, GC21:Sausage\\
&&&&DS23:CTDG-33\\
4&23&44.6\%&$-1.82\pm0.34$&FW22:Group44, FW22:Group3, FW22:Group43, FW22:Group69\\
5&22&67.6\%&$-1.37\pm0.34$&new\\
6&22&66.7\%&$-1.50\pm0.46$&FW22:Group79, DS22a:DTG-2\\
7&19&66.8\%&$-0.61\pm0.32$&DS22b:DTG-2\\
8&19&33.3\%&$-1.66\pm0.26$&GSE, FW22:Group0, SL22:24, FW22:Group5\\
&&&&FW22:Group1, FW22:Group3, DS22a:DTG-7, GC21:Sausage\\
&&&&GM17:Comoving, SM20:Sausage\\
9&18&90.7\%&$-1.36\pm0.20$&GSE, SL22:1, EV21:NGC 6121 (M 4), EV21:VVV CL001\\
10&18&99.6\%&$-0.70\pm0.35$&new\\
\hline
\end{tabular*}
\begin{tablenotes}
      \small
      \item Note—We adopt the nomenclature for previously identified DTGs from \citet{Yuan2020a}. For example, IR18:E is the first and last initial of
the first author (IR) \citep{Roederer2018}, the year the paper was published (2018). Following the colon is the name of the group obtained in that paper (E). 

      \smallskip
Note—We use the following references for associations: GM17: \citet{Myeong2017}, GM18b: \citet{Myeong2018b}, GM18c: \citet{Myeong2018c}, HK18: \citet{Koppelman2018}, HL19: \citet{Li2019}, SM20: \citet{Monty2020}, ZY20a: \citet{Yuan2020a}, EV21: \citet{Vasiliev2021}, DG21: \citet{Gudin2021}, GC21: \citet{Cordoni2021}, GL21: \citet{Limberg2021},  DS22a: \citet{Shank2022a}, DS22b: \citet{Shank2022b}, FW22:\citet{Wang2022}, KH22: \citet{Hattori2022}, KM22: \citet{Malhan2022}, SL22: \citet{Lovdal2022}, DS23: \citet{Shank2023}, JZ23: \citet{Zepeda2023}. 

      \smallskip
      Note—This table is a stub; the full table is available in the electronic edition. 
    \end{tablenotes}
\label{tab:3}
\end{table*}

\begin{table*}
\centering
\caption{Identified DTGs, numbers of stars, and mean cylindrical velocities, cylindrical actions, energies, and eccentricities.}
\begin{tabular*}{0.70\textwidth}{cccccc}
\hline
\hline
Cluster & N stars  & $\left(\langle \text{v}_{r} \rangle, \langle \text{v}_{\phi} \rangle, \langle \text{v}_{z} \rangle \right)$ & $\left(\langle \text{J}_{r} \rangle, \langle \text{J}_{\phi} \rangle, \langle \text{J}_{z} \rangle \right)$ & $\langle E \rangle$ & $\langle \text{ecc}\rangle$  \\ 
&\\
& &  $(\sigma_{\langle \text{v}_{r} \rangle}, \sigma_{\langle \text{v}_{\phi} \rangle}, \sigma_{\langle \text{v}_{z} \rangle})$ & $(\sigma_{\langle \text{J}_{r} \rangle}, \sigma_{\langle \text{J}_{\phi} \rangle}, \sigma_{\langle \text{J}_{z} \rangle})$ & $\sigma_{\langle E \rangle} $ & $\sigma_{\langle \text{ecc}\rangle} $ \\
&\\
& & (km s$^{-1}$) & (kpc km s$^{-1}$) & ($10^{5}$ km$^{2}$ s$^{-2}$) \\
\hline
DTG-1&38&$(-21.3,122.6,-45.8)$&$(437.2,1006.5,995.7)$&$-1.361$&0.495\\
&&(78.9,34.5,138.4)&(159.9,142.3,108.5)&$\phantom{+}0.047$&0.082\\
DTG-2&27&$(15.2,-1.8,-28.2)$&$(995.0,-16.3,395.8)$&$-1.496$&0.968\\
&&(156.7,12.1,103.4)&(66.6,104.3,23.8)&$\phantom{+}0.021$&0.018\\
DTG-3&26&$(-41.2,4.4,15.8)$&$(1006.6,49.1,128.4)$&$-1.566$&0.958\\
&&(153.4,11.0,74.1)&(33.0,95.3,45.7)&$\phantom{+}0.013$&0.025\\
DTG-4&23&$(-18.8,-18.6,97.2)$&$(940.3,-205.0,1865.7)$&$-1.234$&0.641\\
&&(148.9,14.1,139.7)&(248.7,114.1,142.5)&$\phantom{+}0.062$&0.056\\
DTG-5&22&$(23.6,98.7,-19.3)$&$(130.9,409.5,146.1)$&$-1.977$&0.560\\
&&(70.3,19.8,78.7)&(36.5,56.6,26.9)&$\phantom{+}0.021$&0.086\\
DTG-6&22&$(-6.5,185.7,-6.5)$&$(58.6,1063.7,267.8)$&$-1.679$&0.248\\
&&(56.9,36.9,92.6)&(36.6,49.2,26.2)&$\phantom{+}0.021$&0.085\\
DTG-7&19&$(4.7,233.5,-4.1)$&$(13.7,2148.6,8.3)$&$-1.488$&0.091\\
&&(23.7,16.5,17.2)&(14.0,44.5,5.3)&$\phantom{+}0.009$&0.046\\
DTG-8&19&$(-51.6,2.4,-4.7)$&$(1234.5,24.9,210.2)$&$-1.462$&0.962\\
&&(191.3,6.3,107.8)&(59.9,50.7,19.7)&$\phantom{+}0.023$&0.021\\
DTG-9&18&$(-56.7,32.4,-6.0)$&$(424.7,213.7,11.4)$&$-1.931$&0.844\\
&&(17.4,14.1,30.6)&(30.8,78.9,8.1)&$\phantom{+}0.018$&0.045\\
DTG-10&18&$(5.3,236.1,9.6)$&$(18.8,2655.4,8.5)$&$-1.370$&0.104\\
&&(23.2,22.4,17.2)&(30.8,85.5,4.7)&$\phantom{+}0.019$&0.071\\
\hline
\end{tabular*}
\begin{tablenotes}
      \small
      \item Note—This table is a stub; the full table is available in the electronic edition. 
    \end{tablenotes}
\label{tab:4}
\end{table*}

The orbital characteristics of the stars with available RVs are determined using their distances, proper motions, and RVs and corresponding errors as inputs to the Action-based GAlaxy Modelling Architecture\footnote{\url{http://github.com/GalacticDynamics-Oxford/Agama}} (\texttt{AGAMA}) package \citep{Vasiliev2019}, adopting the Solar positions and peculiar motions described in \citet{Shank2022b}\footnote{We adopt a Solar position of ($-$8.249, 0, 0) kpc (\citealt{GRAVITYCollaboration}) and Solar peculiar motion ($U$, $V$ $W$), about the Local Standard of Rest (LSR),  as (11.1,12.24,7.25) km $\text{s}^{-1}$ (\citealt{Schonrich2010}), where $V_\text{LSR} = 238.5 \, \text{km} \, \text{s}^{-1}$, defined as $V_{\text{LSR}} = V_{\odot} - V$ and $V_{\odot} = 250.70$
 $\, \text{km} \, \text{s}^{-1}$, determined from \citet{Reid2020} based on our choice of Solar position and using the proper motion of the centre of the Galaxy (Srg A*) of $-6.411 \, \text{mas} \, \text{yr}^{-1}$.
 We note our main results hold very well when adopting alternative values of Solar peculiar motion and $V_{\rm LSR}$ \citep[e.g.,][]{2015MNRAS.449..162H, 2016MNRAS.463.2623H, 2023ApJ...946...73Z}.}, and the gravitational potential \texttt{MW2017} \citep{McMillan2017}. Similar to \citet{Shank2022b}, the input quantities and their errors are run 1000 times through the orbital-integration process in \texttt{AGAMA} to calculate the  cylindrical velocities (v$_r$, v$_{\phi}$, and v$_z$), 
cylindrical actions (J$_{r}$, J$_{\phi}$, J$_{z}$), orbital specific energy ($E$), and eccentricity (ecc), along with their associated errors. We also determine estimates of the maximum distances from the Galactic plane $Z_{\text{max}}$, the pericentric distance $r_{\text{peri}}$, and apocentric distance $r_{\text{apo}}$. Refer to \citet{Shank2022b} for definitions of these orbital parameters and for the Monte Carlo error calculation treatment. Note that, according to these calculations, all 5355 RRLs in the Final Sample are bound within the Galaxy ($E < 0 \, \text{km}^{2} \, \text{s}^{-2}$). 

\begin{figure*}
\centering
\includegraphics[width=15cm]{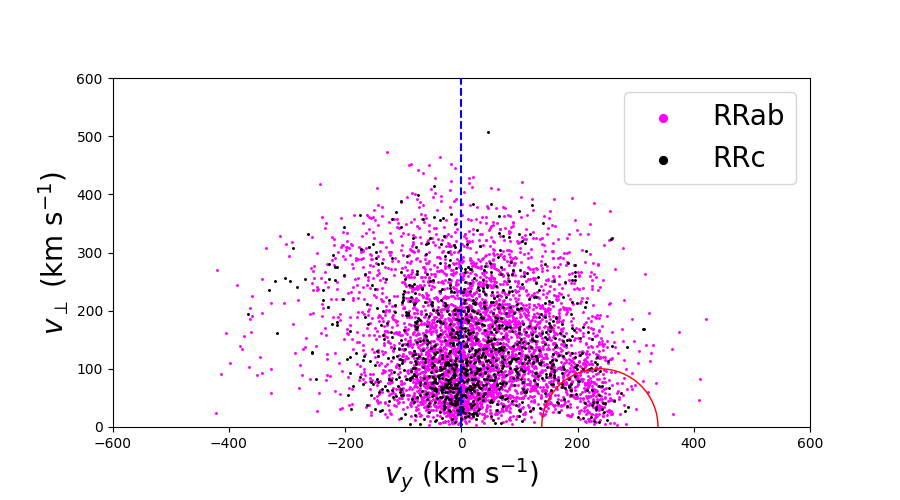}
\caption[Toomre diagram of final RRL sample]
{Toomre Diagram of the Final Sample. The axes are $\text{v}_{\perp} = \sqrt{\text{v}^{2}_{x} + \text{v}^{2}_{z}} $ vs. $\text{v}_{y}$. The red circle represents stars that are within a radius of $100 
\, \text{km} \, \text{s}^{-1}$ from the Local Standard of Rest ($238.5 \, \text{km} \, \text{s}^{-1}$), while the vertical blue dashed line represents the division between prograde ($\text{v}_{y} > 0 \, \text{km} \, \text{s}^{-1}$) and retrograde ($\text{v}_{y} < 0 \, \text{km} \, \text{s}^{-1}$) stellar orbits. For the Final 
Sample of 5355 stars, 58\% and 42\% of RRLs follow prograde and retrograde orbits respectively. There are 4324 RRab stars and 1031 RRc stars that make up the Final 
Sample. For the RRab stars, 59\% and 41\% follow prograde and retrograde orbits, respectively, while the RRc stars follow 56\% and 44\% follow prograde and retrograde orbits, respectively. The fractions of stars within the disk region (red circle) out of number of RRLs in the Final Sample that are RRab and RRc stars are 6.4\%, and 4.5\%, respectively.}
\label{fig:6}
\end{figure*}

A Toomre diagram of the the Final Sample is shown in Fig. \ref{fig:6}. For this sample, 58\% of the RRLs are on prograde orbits, while 42\% are on retrograde orbits. A small fraction of these stars (7\%) fall within a radius of 100 km s$^{-1}$ from the Local Standard of Rest indicated by the red circle in this figure. Of the 5355 RRLs, there are 4324 RRab stars and 1031 RRc stars. For the subset of RRab stars, 59\% follow prograde orbits, while 41\% follow  retrograde orbits, and 6.4\% are within the red circle. For the subset of RRc stars, 56\% follow prograde orbits, while 44\% follow retrograde orbits, and 4.5\% are within the red circle. Thus, the fractions of stars on prograde and retograde orbits for both RRab and RRc stars are roughly the same, justifying our choice to consider them together in our dynamical analysis. 

\begin{figure*}
\centering
\includegraphics[width=16cm]{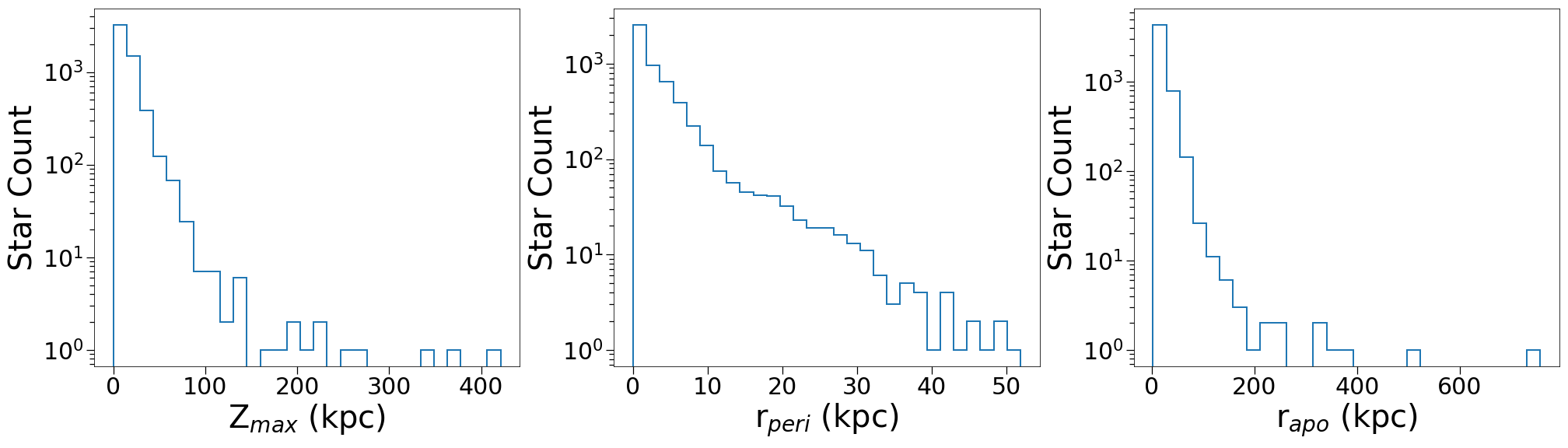}
\caption[Distributions of zmax, rperi, rapo]{Logarithmic histograms for $Z_{\text{max}}$ (left), $r_{\text{peri}}$ (middle), and $r_{\text{apo}}$ (right) for the Final Sample of 5355 RRLs. Note that all stars have $r_{\text{peri}}$ that are within $\sim$ 50 kpc. Most stars fall within $\sim$ 200 kpc for both 
$Z_{\text{max}}$ and $r_{\text{apo}}$. For both these derived quantities, stars beyond $\sim 200$ kpc primarily arise due to large proper-motion errors.}
\label{fig:7}
\end{figure*}

Fig. \ref{fig:7} presents histograms of the derived estimates of $Z_{\text{max}}$, $r_{\text{peri}}$, and $r_{\text{apo}}$. The majority of the stars have their $Z_{\text{max}}$  and $r_{\text{apo}}$ within $\sim$ 200 kpc.  The $r_{\text{peri}}$ distances are all within $\sim$ 50 kpc. There are RRLs with $Z_{\text{max}}$ exceeding 
200 kpc and $r_{\text{apo}}$ with values beyond 200 kpc, which are likely due to errors in these derived quantities primarily arising from the larger input proper-motion errors.

\par

\section{Dynamically Tagged Groups of RR Lyraes in the Milky Way}
\label{sec:DTGs}
The Final Sample of 5355 RRLs with orbital parameters are grouped with the clustering algorithm \texttt{HDBSCAN}, which identifies Dynamically Tagged Groups (DTGs) of stars based on their similar dynamical properties $(E, \text{J}_{r}, \text{J}_{\phi}, \text{J}_{z})$. It is important to note that the metallicities are $not$ used in this procedure.  This algorithm takes into consideration the phase space of the orbital energies and cylindrical actions and uses the density of the data in this space to identify clusters in a hierarchical fashion. For a more detailed description about the \texttt{HDBSCAN} clustering algorithm, we refer the interested reader to \citet{Shank2022b}.

\begin{table*}
\centering
\caption{Identified Milky Way Substructures}
\begin{tabular*}{0.95\textwidth}{ccccccc}
\hline
\hline
MW substructure & N stars & $\langle \text{[Fe/H]} \rangle $ & $\left(\langle \text{v}_{r} \rangle, \langle \text{v}_{\phi} \rangle, \langle \text{v}_{z} \rangle \right)$ & $\left(\langle \text{J}_{r} \rangle, \langle \text{J}_{\phi} \rangle, \langle \text{J}_{z} \rangle \right)$ & $\langle E \rangle$ & $\langle \text{ecc}\rangle$  \\ 
&\\
& & & $(\sigma_{\langle \text{v}_{r} \rangle}, \sigma_{\langle \text{v}_{\phi} \rangle}, \sigma_{\langle \text{v}_{z} \rangle})$ & $(\sigma_{\langle \text{J}_{r} \rangle}, \sigma_{\langle \text{J}_{\phi} \rangle}, \sigma_{\langle \text{J}_{z} \rangle})$ & $\sigma_{\langle E \rangle} $ & $\sigma_{\langle \text{ecc}\rangle} $ \\
&\\
& & & (km s$^{-1}$) & (kpc km s$^{-1}$) & ($10^{5}$ km$^{2}$ s$^{-2}$) \\
\hline
Gaia-Sausage-Enceladus (GSE)
&447&$-1.61$&$(-9.3,-5.3,-2.1)$&$(925.8,-63.2,231.3)$&$-1.581$&0.896\\
&&$\phantom{+}0.31$&(145.0,38.8,81.6)&(422.6,337.4,220.9)&$\phantom{+}0.203$&0.087\\
\hline
Metal-Weak Thick Disk (MWTD)
&70&$-1.35$&$(5.7,131.3,2.7)$&$(171.3,756.3,120.8)$&$-1.821$&0.465\\
&&$\phantom{+}0.46$&(83.8,31.0,70.0)&(88.6,190.2,96.8)&$\phantom{+}0.125$&0.078\\
\hline
Helmi Stream
&65&$-1.78$&$(0.6,134.2,-20.7)$&$(363.2,1073.0,1221.3)$&$-1.335$&0.425\\
&&$\phantom{+}0.33$&(90.9,52.3,124.7)&(174.5,287.3,529.5)&$\phantom{+}0.056$&0.116\\
\hline
Sequoia
&56&$-1.70$&$(-16.7,-178.8,66.0)$&$(723.3,-1698.0,848.0)$&$-1.226$&0.522\\
&&$\phantom{+}0.26$&(107.1,91.2,119.5)&(371.3,732.3,583.8)&$\phantom{+}0.079$&0.133\\
\hline
Sagittarius
&34&$-1.89$&$(-59.9,100.4,-99.9)$&$(1020.9,1428.8,3067.0)$&$-1.026$&0.464\\
&&$\phantom{+}0.46$&(135.5,38.9,151.7)&(662.2,562.8,1281.6)&$\phantom{+}0.217$&0.097\\
\hline
LMS-1 (Wukong)
&8&$-2.14$&$(-4.4,47.3,108.6)$&$(448.1,361.2,1495.1)$&$-1.387$&0.526\\
&&$\phantom{+}0.15$&(150.4,17.5,107.7)&(88.1,98.9,87.3)&$\phantom{+}0.021$&0.051\\
\hline
\end{tabular*}
\label{tab:5}
\end{table*}

\begin{table*}
\caption{Associations of Identified DTGs}
\begin{center}
\begin{tabular}{ cccc }
\hline
\hline
Structure & Reference & Associations &Identified DTGs \\
\hline
\multirow{8}{8em}{MW Substructure} & \multirow{8}{8em}{\citealt{Naidu2020}} &Gaia-Sausage-Enceladus (GSE)
&2, 3, 8, 9, 11, 13, 14, 15, 16, 20, 21, 23, 29, 30 \\
&&&33, 36, 38, 40, 45, 48, 50, 51, 53, 56, 62, 64, 65, 66\\
&&&68, 69, 70, 72, 73, 76, 77, 80, 82, 84, 86, 87, 88, 89\\
&&&90, 91\\\cline{3-4}
&&Metal-Weak Thick Disk (MWTD)
&12, 44, 58, 61, 63, 71, 75, 85, 97\\\cline{3-4}
&&Helmi Stream
&1, 22, 25\\\cline{3-4}
&&Sequoia
&18, 24, 27, 57, 95\\\cline{3-4}
&&Sagittarius
&28, 37, 79, 92\\\cline{3-4}
&&LMS-1 (Wukong)
&52\\
\hline
\multirow{6}{10em}{Globular Clusters} & \multirow{6}{14em}{\citealt{Vasiliev2021}} &Ryu 879 (RLGC 2)&11, 30, 50\\\cline{3-4}
&&VVV CL001&24\\\cline{3-4}
&&Pal 5&9\\\cline{3-4}
&&Terzan 10&17\\\cline{3-4}
&&NGC 6388&20\\\cline{3-4}
&&IC 1257&39\\
\hline
\end{tabular}
\end{center}
\label{tab:6}
\end{table*}


The following inputs are set to initiate \texttt{HDBSCAN}: \texttt{min\_cluster\_size} = 5, \texttt{min\_samples} = 5, \texttt{cluster\_selection\_method} = \texttt{'leaf'}, \texttt{prediction\_data} = \texttt{'True'}, Monte Carlo runs set to 1000, and the minimum confidence level is set to 20\% for a cluster to pass the confidence test. The \texttt{min\_cluster\_size} sets the minimum number of stars allowed in a cluster, and the \texttt{min\_samples} is set to the \texttt{min\_cluster\_size}, which is used to account for noise levels in the data. Setting \texttt{cluster\_selection\_method} = \texttt{'leaf'} permits tighter clustering, and \texttt{prediction\_data} = \texttt{'True'} allows \texttt{HDBSCAN} to store memory of nominal clusters for the subsequent Monte Carlo run. We found that setting the \texttt{min\_cluster\_size} input to 5, as opposed to a higher number such as 10, has little impact on the mean   confidence levels associated with the identified DTGs, while providing a superior resolution of the fine-scale dynamical substructure. 

The \texttt{HDBSCAN} clustering routine found 97 DTGs. Table \ref{tab:3} lists the identified DTGs and their corresponding number of stellar members, the confidence levels, the means and dispersions of their metallicities, and associations made with MW substructures, stellar associations, previous group associations, GC associations, and dwarf galaxy associations. Note that the nomenclature for previous group associations for DTGs is adopted from \citet{Yuan2020a}. If a DTG is not associated with any of the types of association, it is labeled with the word `new'. The average confidence level for the 97 DTGs is 48.6\%.  The average confidence level for the DTGs with 10 or more members is 50.1\%; for the DTGs with fewer than 10 members, the average confidence level is 45.1\%.

Table \ref{tab:7} lists the DTGs identified by \texttt{HDBSCAN} with their associations and respective references, along with the names (using $Gaia$ IDs) and [Fe/H] for each member of the particular DTG. 

Table \ref{tab:4} lists the mean cluster dynamical parameters for each DTG, their corresponding number of stellar members, along with mean cylindrical velocities, cylindrical actions, energies, eccentricities, and their corresponding dispersions. 

We emphasize that the DTGs of RRLs we identify can also be used to target non-variable stars with similar dynamical parameters for which more complete elemental-abundance studies can be carried out in the future.

\subsection{Milky Way Substructures}
\label{subsec:DTGs_sub1}
The DTGs are associated with known MW substructures. These associations are determined by the orbital parameters and [Fe/H] for the average of the individual RRLs present in each DTG. If the DTG meets the criteria of a particular substructure, then they are associated with that substructure. We generally adopt the criteria presented in \citet{Naidu2020} to assign associations with MW substructures. As a result, we identify DTGs associated with MW substructures such as the Gaia-Sausage-Enceladus (GSE), Metal-Weak Thick Disk (MWTD), Helmi Stream, Sequoia, Sagittarius Stream, and LMS-1 (Wukong). However, since the current final sample of RRLs with orbital dynamical parameters do not have $[\alpha/\text{Fe}]$ ratios available, we remove the Naidu criteria associated with $[\alpha/\text{Fe}]$ ratios. This influences how we treat the criteria for the MWTD, so we employ the global rotational velocity range quoted in \citet{Carollo2010} as a criterion. 

Table \ref{tab:5} lists the MW substructures that are identified as associations, their corresponding  number of stellar members, their means and dispersions of metallicity, and the mean cylindrical velocities, cylindrical actions, energies, and eccentricities. 

Table \ref{tab:6} lists the DTGs associated with each identified substructure mentioned in Table \ref{tab:5}. 

The most populated substructure is GSE with 447 members. The second-most populated substructure is the MWTD with 70 members. The third-most populated substructure is the Helmi Stream with 65 members. The fourth-most populated substructure is Sequoia with 56 members. The fifth-most populated substructure is the Sgr Stream with 34 members. LMS-1 (Wukong) is the least-populated substructure, with 8 members. 

\begin{figure*}
\centering
\includegraphics[width=12cm]{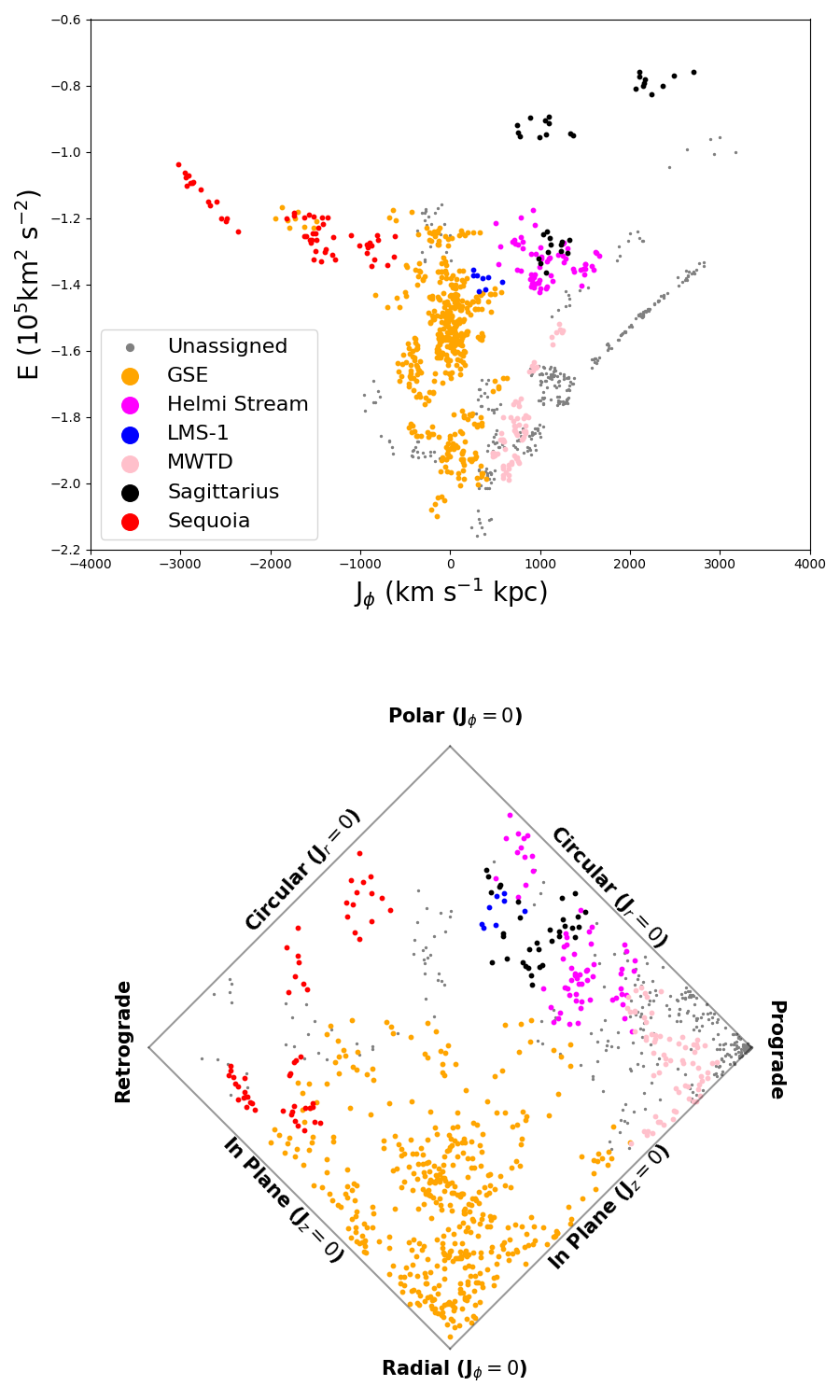}
\caption[Lindblad Diagram for MW substructures]{Top panel: Lindblad Diagram of the identified MW substructures. The different substructures are associated with the different colours in the legend. Bottom Panel: The projected-action plot of the same substructures. This space is represented by $J_{\phi}$/$J_{\text{Tot}}$ on the horizontal axis and $\left(J_{z}- J_{r}\right)$/$J_{\text{Tot}}$ on the vertical axis with $J_{\text{Tot}} = J_{r} + |J_{\phi}| + J_{z}$. For more details on the projected-action space diagram, see Figure 3.25 in \citet{Binney2008}.}
\label{fig:8}
\end{figure*}

Fig. \ref{fig:8} shows the Lindblad Diagram (top panel) and projected-action plot (bottom panel) for the MW substructures listed in Table \ref{tab:6}. The Lindblad Diagram describes the energy and azimuthal-action positions (E, $\text{J}_{\phi}$) of the respective DTG members. The projected-action plot captures the contributions of the cylindrical radial, azimuthal, and vertical components ($\text{J}_{r}$, $\text{J}_{\phi}$, $\text{J}_{z}$) of the actions. A detailed description of the meaning of null and non-zero actions ($\text{J}_{i} = 0$ , $\text{J}_{i} \neq 0$) is provided in \citet{Shank2022b}. 

The GSE substructure exhibits the widest range in energies and the most extended radial component compared to the other MW substructures. GSE members also show a null azmimuthal component, with the exception of a handful of DTGs, with DTG-40 overlapping in action space with Sequoia. This DTG is associated with GSE instead of the Helmi Stream due to its high average eccentricity ($\langle ecc \rangle > 0.7$) and average [Fe/H] not meeting one of the criteria for the 
Helmi Stream ($\langle \text{[Fe/H]} \rangle < -1.6$).   Also, a few GSE members are on planar orbits. Helmi Stream members are all prograde, have a stronger vertical component compared to their radial components, and exhibit more circular orbits. The LMS-1 (Wukong)
substructure exhibits a prograde behaviour with a significant vertical component. The Sgr Stream has three distinct positions, one coinciding in the action-space plot with Helmi Stream members; the two others are isolated from the rest of the DTGs in the Lindblad Diagram with the highest energies. They also have more vertical contributions in their orbits, with a handful close to circular in the projected-action space. The MWTD exhibits prograde orbits with roughly equal vertical and radial components, with the exception of some DTGs being more planar in their orbits than circular and vice versa. Finally, Sequoia members show retrograde orbits, and overlap with one DTG of GSE (DTG-40); two of its four DTGs are more planar, while the remaining two have more circular 
orbits. 

It is interesting to note that the large number of DTGs (43) associated with the GSE substructure in this work was actually presaged by the analysis of \citet{Schlaufman2009}, \citet{Schlaufman2011}, and \citet{Schlaufman2012}, who considered Elements of Cold HalO Substructure (ECHOS) in the inner-halo region based on over-densities in radial-velocity space along SDSS/SEGUE plug-plate lines-of-site.  

In the previous work of \citet{Shank2023}, the MWTD and Splashed Disk substructures shared CDTGs overlapping in the energy-action space due to the criteria selected quoted in \citet{Naidu2020}, which not only utilizes a metallicity range, but also employs a range in [$\alpha$/\text{Fe}] ratios. Due to the absence of alpha-to-iron ratios in our RRL sample, we replace the [$\alpha$/\text{Fe}] criterion in \citet{Naidu2020} with a global rotational velocity criterion $100 < \langle V_{\phi} \rangle < 150 \, \text{km} \,\text{s}^{-1}$ for the MWTD quoted in \citet{Carollo2010}.
For the Splashed Disk, we employed a velocity range $25 < \langle V_{\phi} \rangle < 175 \, \text{km} \,\text{s}^{-1}$ and a metallicity range $-1.0 < \langle \text{[Fe/H]} \rangle < -0.2 \, $, adopted from  \citet{An2021}. The limited numbers of of metal-rich RRLs in our sample precluded association with the Splashed Disk.

\section{Chemical Structure of the Identified RRL DTGs}\label{sec:stats}

\subsection{Statistical Framework}
\label{subsec:stats_sub1}

Following \citet{Gudin2021}, we perform a statistical analysis of our DTGs to determine how probable the observed abundance dispersions of [Fe/H] would be if their member stars were selected at random from the full set of RRL stars in the Final Sample. To perform this analysis, we create $2.5\times 10^6$ random groups of $5 \leq N \leq 38$ stars with dispersions based on the biweight scale \citep{Beers1990}. We then use this to generate cumulative distribution functions (CDFs) for the [Fe/H] distributions for each possible size of the random groups.  If, for a given element, the DTGs preferentially populate the low end of the simulated CDFs, we infer that its members exhibit strong similarities in their metallicities. We then use binomial and multinominal probabilities to calculate the statistical likelihoods of the distributions of the DTG abundance dispersions based on the derived CDFs.

We calculate the statistical significance at three thresholds of the CDF values for [Fe/H] ($\alpha \in \{0.50, 0.33, 0.25\}$), as well the significance across all $\alpha$ values, resulting in two probabilities:  (i) The \textit{Individual Elemental-Abundance Dispersion (IEAD) probability}, which is the individual binomial probability for specific values of $\alpha$ and [Fe/H], and (ii) 
The \textit{Global Element Abundance Dispersion (GEAD) probability}, which is  the multinomial probability for [Fe/H], grouped over all values of $\alpha$. This represents the overall statistical significance.  

For a more detailed discussion of the above probabilities, and their use, the interested reader is referred to \citet{Gudin2021}.

\subsection{Results}
\label{subsec:stats_sub2}

The number of the 97 DTGs satisfying the $\alpha$ levels of 0.50, 0.33, and 0.25 obtained by our calculations are 66, 54, and 47, respectively, which are individually highly statistically significant, as their IEAD probabilities are each $p << 0.001$. We underscore that almost half of the sample of DTGs lies below the $\alpha = 0.25$ level. Unsurprisingly, the GEAD probability (which considers all three levels of $\alpha$) is also highly significant, $p << 0.001$.  We conclude that the stars in the DTGs, which we identify on the basis of their dynamical parameters alone, share common star-formation and chemical histories, influenced by their birth environments. 

\section{Summary and Future Prospects}
\label{sec:summary}

We have analysed a large sample of 135,873 RR Lyrae stars (RRLs), identified on the basis of variations in light curves from $Gaia$ DR3, and with precise photometric-metallicity and distance estimates from the newly calibrated 
$P$--$\phi_{31}$--$R_{21}$--[Fe/H] and $Gaia$ $G$-band $P$--$R_{21}$--[Fe/H]  absolute magnitude-metallicity relations of \citet{Li2023}, combined with available proper motions from $Gaia$ EDR3, and 6955 systemic radial velocities from $Gaia$ DR3 and other sources, in order to explore the chemistry and kinematics of the halo of the MW.

After removal of stars that are potential members of globular clusters and dwarf galaxies, the Magellanic Clouds, and the Sagittarius Stream, and excision of possible artefacts, we obtain a Cleaned Sample of 78,898 RRLs, 78,740 of which fall in the metallicity range $-4.0 <$ [Fe/H] $ \leq +0.5$. 

We use the subset of 42,195 stars in this sample located at $|Z| > 3$ kpc to consider the nature of the MDFs for MW halo RRLs as functions of Galactocentric distance and distance from the Galactic plane (centered on the Sun). By application of a Gaussian Mixture Model, under the assumption that the halo system comprises two stellar populations (the dual-halo interpretation advocated by \citealt{Carollo2007,Carollo2010}, and \citealt{Beers2012}), we demonstrate that this simple model can account for the observed \textit{in-situ} distribution of [Fe/H] remarkably well.  Some small deviations remain between the model prediction and the observations, which may arise from imperfect removal of recognized substructures and artefacts.  

Specializing to the subset of 5355 RRLs in the Cleaned Sample with available RVs in the metallicity range $-4.0 < $ [Fe/H] $ \leq +0.5$, we estimate their dynamical parameters in an assumed static gravitational
potential of the MW.  We apply the \texttt{HDBSCAN} clustering method to the specific energies and cylindrical actions (E, J$_{r}$, J$_{\phi}$, J$_{z}$), identifying 97 Dynamically Tagged Groups (DTGs) of RRLs, and explore their associations with recognized substructures of the MW: $Gaia$-Sausage-Enceladus (447  stars / 44 DTGs), the Metal-Weak Thick Disk (70 stars / 9 DTGs), the Helmi Stream (65 stars / 3 DTGs), Sequoia (56 stars / 5 DTGs), the Sagittarius Stream (34 stars / 4 DTGs), and LMS-1 (Wukong) (8 stars / 1 DTG). 
There are six associations with globular clusters: NGC 6388,  Terzan 10, Pal 5, Ryu 879, IC 1257, and VVV~CL001.  There are 56 DTGs associated with dynamical groups identified in previous works, and 31 DTGs that are newly identified. 

The Lindblad Diagram and the projected-action space of the MW substructures reveal that GSE indeed has the most spread in its energy profile and a strong vertical component. The substructures of Helmi Stream, LMS-1 (Wukong), MWTD, and Sgr Stream are all prograde and have more circular orbits.  The MWTD 
is the most prograde and most metal-rich substructure identified.  There are no DTGs associated with the Splashed Disk, as expected, due to the dearth of metal-rich RRLs. 
Sequoia members are all on retrograde orbits, while there is an interesting split between members that have stronger vertical components compared to their horizontal components, and vice versa, a behaviour not seen in the other substructures. 

An analysis of the dispersions of [Fe/H] for the RRL DTGs yields highly statistically significant (low) dispersions of [Fe/H] for the stellar members of the DTGs compared to random draws from the full 
sample, indicating that they share common star-formation and chemical histories, influenced by their birth environments. 

To our knowledge, the MW RRL sample we investigate is the largest to date. It is already yielding important constraints on the nature of the halo system, its substructures, and its formation and evolution.  
In Paper VI of this series, we plan to further explore the nature of the MW halo system, and make use of the sub-sample of RRLs with available RVs in order to construct measures of its global structure, providing strong constraints on the shape of the dark matter halo.

We anticipate that a much larger sample of RRLs will become available with the next data release from $Gaia$, anticipated within the next two years.  This will not only enable improved estimates of photometric metallicities and distances, as well as provide additional systemic radial velocities, but will expand the number and reach of the sample suitable for \textit{in-situ} studies of the halo system.

\section*{Acknowledgements}
\label{sec:Acknowledgements}

J.C.G., T.C.B., J.H., D.S., D.S., D.G., and D.K. acknowledge partial support for this work from grant PHY 14-30152; Physics Frontier Center/JINA Center for the Evolution of the Elements (JINA-CEE), and OISE-1927130: The International Research Network for Nuclear Astrophysics (IReNA), awarded by the US National Science Foundation. 
H.W.Z. acknowledge support from the National Key R\&D Program of China (No. 2019YFA0405500), the National Natural Science Foundation of China (NSFC grant Nos. 11973001, 12090040, and 12090044), and the science research grants from the China Manned Space Project with No. CMS-CSST-2021-B05.
G.C.L. acknowledge support from the key Laboratory Fund of Ministry of Education under grant No. QLPL2022P01 and NSFC grant No. U1731108.
Y.S.L. acknowledges support from the National
Research Foundation (NRF) of Korea grant funded by the Ministry of Science and ICT (NRF-2021R1A2C1008679) Y.S.L. also acknowledges partial support for his visit to the University of Notre Dame from OISE-1927130: The International Research Network for Nuclear Astrophysics (IReNA), awarded by the US National Science Foundation. 
Y.H. was supported by JSPS KAKENHI Grant Numbers JP22KJ0157, JP20K14532, JP21H04499, JP21K03614, JP22H01259.

This work has made use of data from the European Space Agency (ESA) mission $Gaia$ (https://www.cosmos.esa.int/gaia), processed by the $Gaia$ Data Processing and Analysis Consortium (DPAC, https://www.cosmos.esa.int/web/gaia/dpac/consortium). 

\section*{Appendix}
Here we present the table for the DTGs identified by \texttt{HDBSCAN} (Table \ref{tab:7}). The full table is available as an electronic version. Note that the full tables for Table \ref{tab:3} and \ref{tab:4} are also available in their respective electronic versions. 

\begin{table*}
\centering
\caption{DTGs Identified by \texttt{HDBSCAN}}
\begin{tabular*}{0.65\textwidth}{cccccc}
\hline
\hline
Star Name & [Fe/H]\\
\hline
\multicolumn{2}{c}{DTG-1}\\
\multicolumn{2}{c}{Structure: Helmi Stream
}\\
\multicolumn{2}{c}{Group Assoc: (CDTG-15: \citealt{Gudin2021})}\\
\multicolumn{2}{c}{Group Assoc: (DTG-3: \citealt{Limberg2021b})}\\
\multicolumn{2}{c}{Group Assoc: (DTG-42: \citealt{Shank2022b})}\\
\multicolumn{2}{c}{Stellar Assoc: 1385 (Group69: \citealt{Wang2022})}\\
\multicolumn{2}{c}{Stellar Assoc: 1573 (Group71: \citealt{Wang2022})}\\
\multicolumn{2}{c}{Stellar Assoc: 1764 (Group73: \citealt{Wang2022})}\\
\multicolumn{2}{c}{Stellar Assoc: 244 (Group69: \citealt{Wang2022})}\\
\multicolumn{2}{c}{Stellar Assoc: 250 (Group69: \citealt{Wang2022})}\\
\multicolumn{2}{c}{Stellar Assoc: 5711 (DTG-3: \citealt{Yuan2020a})}\\
\multicolumn{2}{c}{Stellar Assoc: 736 (Group73: \citealt{Wang2022})}\\
\multicolumn{2}{c}{Stellar Assoc: 1454724187872808832 (NGC 5272 (M 3): \citealt{Vasiliev2021})}\\
\multicolumn{2}{c}{Stellar Assoc: 2174 (Group75: \citealt{Wang2022})}\\
\multicolumn{2}{c}{Stellar Assoc: 1685 (Group71: \citealt{Wang2022})}\\
\multicolumn{2}{c}{Stellar Assoc: 888 (Group74: \citealt{Wang2022})}\\
\multicolumn{2}{c}{Stellar Assoc: 1443 (Group74: \citealt{Wang2022})}\\
\multicolumn{2}{c}{Stellar Assoc: 1669 (Group70: \citealt{Wang2022})}\\
\multicolumn{2}{c}{Stellar Assoc: 1147 (Group73: \citealt{Wang2022})}\\
\multicolumn{2}{c}{Stellar Assoc: 2601 (Group73: \citealt{Wang2022})}\\
\multicolumn{2}{c}{Stellar Assoc: 1996 (Group70: \citealt{Wang2022})}\\
\multicolumn{2}{c}{Stellar Assoc: 1127 (Group72: \citealt{Wang2022})}\\
\multicolumn{2}{c}{Stellar Assoc: 1411 (Group74: \citealt{Wang2022})}\\
\multicolumn{2}{c}{Globular Assoc: No Globular Associations 
}\\
\multicolumn{2}{c}{Dwarf Galaxy Assoc: No Dwarf Galaxy Associations 
}\\
\hline
4459079134549605504&$-1.61$\\
3187615948455925120&$-1.98$\\
3452489978919442176&$-1.36$\\
3951110771173681024&$-1.45$\\
3958056665299734144&$-0.73$\\
37911886777525888&$-1.22$\\
1269371644394005120&$-1.74$\\
379490807626518016&$-1.85$\\
739754068867610112&$-1.33$\\
3698088617065361792&$-1.72$\\
3699931089316617728&$-2.32$\\
3701890487755176704&$-1.64$\\
3704536703005861504&$-1.78$\\
2688868643643546880&$-1.25$\\
4450058431917932672&$-1.70$\\
4661506200297885184&$-1.20$\\
56854857216208512&$-1.19$\\
1009665142487836032&$-1.50$\\
1022285989786150656&$-2.10$\\
1453588392356674816&$-1.74$\\
1454724187872808832&$-1.90$\\
1454777892144186624&$-1.83$\\
1454780709642463744&$-2.16$\\
1454783149183752576&$-1.73$\\
1454874717884070400&$-1.49$\\
1454879219009819136&$-2.07$\\
1454798782864780288&$-1.66$\\
1454881314953875840&$-1.68$\\
1454892172631004032&$-1.66$\\
5825290639263641472&$-1.59$\\
1480652905435076480&$-2.39$\\
4009616510737850112&$-1.57$\\
4013370788895170048&$-1.91$\\
4564842429335049856&$-1.60$\\
1264657655792671104&$-1.94$\\
1301520952074594048&$-1.68$\\
5248858577304291328&$-2.08$\\
1347461017487423104&$-1.41$\\
$\mu$ $\pm$ $\sigma$ \, $\left(\text{[Fe/H]}\right)$&$-1.69\pm0.33$\\
\hline
\end{tabular*}
\begin{tablenotes}
      \small
      \item Note—This table is a stub; the full table is available in the electronic edition. 
\end{tablenotes}
\label{tab:7}
\end{table*}


\clearpage
\bibliographystyle{mnras}
\DeclareRobustCommand{\DE}[3]{#3}
\bibliography{main}

\bsp	
\label{lastpage}
\end{document}